\documentstyle[12pt,A4]{article}
\catcode`\@=11

\@addtoreset{equation}{section}
\def\theequation{\arabic{section}.\arabic{equation}}

\def\@normalsize{\@setsize\normalsize{15pt}\xiipt\@xiipt
\abovedisplayskip 14pt plus3pt minus3pt%
\belowdisplayskip \abovedisplayskip
\abovedisplayshortskip  \z@ plus3pt%
\belowdisplayshortskip  7pt plus3.5pt minus0pt}

\def\small{\@setsize\small{13.6pt}\xipt\@xipt
\abovedisplayskip 13pt plus3pt minus3pt%
\belowdisplayskip \abovedisplayskip
\abovedisplayshortskip  \z@ plus3pt%
\belowdisplayshortskip  7pt plus3.5pt minus0pt
\def\@listi{\parsep 4.5pt plus 2pt minus 1pt
            \itemsep \parsep
            \topsep 9pt plus 3pt minus 3pt}}

\def\underline#1{\relax\ifmmode\@@underline#1\else
        $\@@underline{\hbox{#1}}$\relax\fi}
\@twosidetrue

\relax

\catcode`@=12

\catcode`\@=11

\def\section{\@startsection{section}{1}{\z@}{3.5ex plus 1ex minus
   .2ex}{2.3ex plus .2ex}{\large\bf}}

\def\thesection{\Roman{section}.}

\def\appendix{\setcounter{section}{0}
        \def\thesection{APPENDIX }
        \def\theequation{\Alph{section}.\arabic{equation}}}

\def\FERMIPUB{}

\def\ps@headings{\def\@oddfoot{}\def\@evenfoot{}
\def\@oddhead{\hbox{}\hfill
        \makebox[.5\textwidth]{\raggedright\ignorespaces --\thepage{}--
        \hfill {\rm FERMILAB--Pub--\FERMIPUB}}}
\def\@evenhead{\@oddhead}
\def\subsectionmark##1{\markboth{##1}{}}
}

\catcode`\@=12

\relax

\def\figcap{\section*{Figure Captions\markboth
        {FIGURECAPTIONS}{FIGURECAPTIONS}}\list
        {Fig. \arabic{enumi}:\hfill}{\settowidth\labelwidth{Fig. 999:}
        \leftmargin\labelwidth
        \advance\leftmargin\labelsep\usecounter{enumi}}}
 \relax
\def\tablecap{\section*{Table Captions\markboth
        {TABLECAPTIONS}{TABLECAPTIONS}}\list
        {Table \arabic{enumi}:\hfill}{\settowidth\labelwidth{Table 999:}
        \leftmargin\labelwidth
        \advance\leftmargin\labelsep\usecounter{enumi}}}
 \relax
\def\reflist{\section*{References\markboth
        {REFLIST}{REFLIST}}\list
        {[\arabic{enumi}]\hfill}{\settowidth\labelwidth{[999]}
        \leftmargin\labelwidth
        \advance\leftmargin\labelsep\usecounter{enumi}}}
 \relax

\catcode`\@=11

\def\FERMIPUB{}

\def\ps@headings{\def\@oddfoot{}\def\@evenfoot{}
\def\@oddhead{\hbox{}\hfill
        \makebox[.5\textwidth]{\raggedright\ignorespaces --\thepage{}--
        \hfill {\rm FERMILAB--Pub--\FERMIPUB}}}
\def\@evenhead{\@oddhead}
\def\subsectionmark##1{\markboth{##1}{}}
}

\ps@headings

\relax
\newskip\humongous \humongous=0pt plus 1000pt minus 1000pt

\newif\ifdtup

\def\beq{\begin{equation}}
\def\eeq{\end{equation}}

\def\beqn{\begin{eqnarray}}
\def\eeqn{\end{eqnarray}}
\relax

\def\G2{{\; \rm GeV/}c^2}
\def\G{\; \rm GeV}

\def\dotx{\dotx{\dot\overline{x}}}

\relax

\begin{document}
\hbadness=10000
\begin{titlepage}
\nopagebreak
\begin{flushright}

        {\normalsize
 Kanazawa-93-12\\
 KITP-9303\\

November, 1993   }\\
\end{flushright}
\vspace{2.5 cm}
\vfill
\begin{center}
\renewcommand{\thefootnote}{\fnsymbol{footnote}}
{\large \bf Levels of U(1)$_Y$ in Minimal String Model\\
on $Z_N\times Z_M$ Orbifolds }
\vfill
{\bf Hiroshi Kawabe, Tatsuo Kobayashi } and

{\bf Noriyasu Ohtsubo$^*$}\\

\vspace{1.5cm}
       Department of Physics, Kanazawa University, \\
       Kanazawa, 920-11, Japan \\
 and \\
$^*$Kanazawa Institute of Technology, \\
Ishikawa 921, Japan

\vfill
\end{center}

\vfill
\nopagebreak
\begin{abstract}
We study a minimal string model possessing the same massless spectra as the
 MSSM on $Z_N\times Z_M$ orbifolds.
Threshold corrections of the gauge coupling constants of SU(3), SU(2) and
 U(1)$_Y$ are investigated in a case of an overall modulus.
Using computer analyses, we search ranges of levels of U(1)$_Y$ allowed by
 the LEP experiments.
It is found that $Z_3\times Z_3$ can not derive the minimal string model for
 a $M_Z$ SUSY breaking scale.
The minimum values of the overall moduli are estimated within the ranges of
 the levels.

\end{abstract}

\vfill
\end{titlepage}
\pagestyle{plain}
\newpage
\voffset = 0 cm

\leftline{\large \bf 1. Introduction}
\vspace{0.8 cm}

An orbifold compactification is one of the most attractive procedures deriving
 a 4-dim unified theory from superstring theories \cite{ZNOrbi}.
Much work has been devoted in order to construct the minimal supersymmetric
 standard model (MSSM) from the orbifold compactification and to study their
 phenomenological aspects \cite{ZNOrbi2,Anom}.
The superstring theories imply that even without a unification of gauge
 groups, gauge coupling constants are unified at a string scale
$M_{\rm string}=5.27\times g_{\rm string}\times 10^{17}$ GeV
 \cite{Kaplunovsky,Derendinger}, where $g_{\rm string}\simeq 1/\sqrt 2$ is
 the universal string coupling constant.
On the other hand, recent LEP measurements support that gauge coupling
 constants  of the standard gauge group ${\rm SU(3)\times SU(2)\times U(1)}_Y$
 are unified at $M_X\simeq 10^{16}$ GeV within the framework of the MSSM
 \cite{MSSM} with a Kac-Moody level $k_1=5/3$ of ${\rm U(1)}_Y$ proposed by
 GUTs.
It seems that this difference between $M_X$ and $M_{\rm string}$
 rejects the possibility for a minimal string model which has the same massless
 spectra as the MSSM.
The minimal string model may include hidden gauge groups, extra U(1)'s and
 extra pairs of $(3,2)+(\bar 3,2)$, etc., where the extra U(1)'s and the extra
 pairs must be removed through some breaking mechanisms such as the anomalous
 U(1) breaking \cite{Anom}.
The hidden gauge groups are expected to contribute to the SUSY breaking.

It has been pointed out that the difference of the two mass scales may be
 explained by threshold effects due to towers of higher massive modes.
The threshold corrections have been calculated in the case of the orbifold
 models \cite{Kaplunovsky,Derendinger,Dixon}.
In the calculation a target-space duality symmetry \cite{Kikkawa} plays an
 important role.
In ref.~\cite{Ibanez}, the unification of the gauge coupling constants of
 SU(3), SU(2) and U(1)$_Y$ was studied with considering the threshold effects
 in the case of $k_1=5/3$.
It was shown that the minimal string model may be derived from $Z_6$-II,
 $Z_8$-I and the whole $Z_N\times Z_M$ (except fot $Z_2\times Z_2$) orbifold
 models \cite{ZNM,KO} without
 conflicting the unification of the coupling constants.
For the $Z_8$-I, an explicit research for the minimal string model was studied
 without such a restriction of $k_1$ in ref.~\cite{KKO}.
There is no reason why we choose $k_1=5/3$ in the minimal string model, where
 the level is arbitrary.

In this paper, we study the minimal string model on $Z_N\times Z_M$ orbifolds
 and derive ranges of the levels which lead threshold corrections consistent
 with measured values of the gauge coupling constants.
In the next section, we briefly review the $Z_N\times Z_M$ orbifold models and
 discuss massless conditions to constrain oscillator numbers and the levels.
In section three, we review the duality symmetry and the threshold corrections
 to the gauge coupling constants in a case of an overall modulus.
Then we study the unification of the SU(3) and SU(2) gauge coupling constants
 in the minimal string model derived from the $Z_N\times Z_M$ orbifold models.
In section four, we estimate the ranges of the allowable levels of U(1)$_Y$
 through computer analyses and then we get minimum values of the overall
 moduli.
The last section is devoted to the conclusions.

\vspace{0.8 cm}
\leftline{\large \bf 2. $Z_N\times Z_M$ Orbifold Models}
\vspace{0.8 cm}

In the orbifold models, the string states consist of the bosonic strings on the
 4-dim space-time and a 6-dim orbifold, their right-moving superpartners and
 left-moving gauge parts whose momenta span a shifted $E_8 \times E'_8$
 lattice.
The right-moving fermionic parts are bosonized and momenta of the bosonized
 fields span an SO(10) lattice.
The $Z_N\times Z_M$ orbifolds are obtained through the division of 6-dim tori
 by twists $\theta$ and $\omega$ of order $N$ and $M$ respectively.
We denote eigenvalues of the twists $\theta$ and $\omega$ in a complex basis
 ($X_i,\tilde X_i$) ($i=1,2,3$) as exp$[2\pi i v^i_1]$ and exp$[2\pi i v^i_2]$
 respectively, whose exponents $v_1$ and $v_2$ are exibited in the second
 column of Table 1.
The twists $\theta$ and $\omega$ are embedded into the SO(10) and
 $E_8 \times E'_8$ lattices in terms of shifts so that the $N=1$ SUSY remains
 and the gauge group breaks into a small one.
The $E_8 \times E'_8$ lattice is shifted by Wilson lines \cite{WL}, as well.

There are two types of closed strings on the orbifolds.
One is an untwisted string whose massless states should satisfy
$$h-1=0, \eqno(2.1)$$
where $h$ is a conformal dimension of the $E_8\times E'_8$ gauge part.
The other is a twisted string.
Massless states of $\theta^\ell \omega^m$-twisted sector $T_{\ell m}$
 \cite{KO} should satisfy the following condition:
$$h+N_{\rm OSC}+c_{\ell m}-1=0, \eqno(2.2)$$
where $N_{\rm OSC}$ is an oscillation number and $c_{\ell m}$ is obtained from
$$ c_{\ell m}= {1\over 2}\sum_{i=1}^3 v^i_{\ell m}(1-v^i_{\ell m}),$$
$$v^i_{\ell m} \equiv \ell v^i_1+mv^i_2-{\rm Int}(\ell v^i_1+mv^i_2).
\eqno(2.3)$$
Here ${\rm Int}(a)$ represents an integer part of $a$.

A representation $\underline{R}$ of the non-abelian group $G$ contributes to
 the conformal dimension as
$$ h={C(\underline{R}) \over C(G)+k},\eqno(2.4)$$
where $k$ is a level of a Kac-Moody algebra corresponding to $G$ and
 $C(\underline{R})$ ($C(G)$) is a quadratic Casimir of the $\underline{R}$
 (adjoint) representation, e.g., $C({\rm SU}(N))=N$.
In general the string theories derive the gauge groups with $k=1$,
 except for U(1).
Then we restrict ourselves to the case where $k=1$ for SU(3) and SU(2).
In the minimal string model, the level $k_1$ of U(1)$_Y$ is a
 non-negative free parameter.

The representations (3,2), ($\overline 3,1$) and (1,2) of the
 ${\rm SU(3) \times SU(2)}$ group contribute to the conformal dimension as
 $h=7/12,$ 1/3 and 1/4, respectively.
A state with a charge $Q$ of the U(1)$_Y$ group gives $h=Q^2/k_1$.
{}From eq.~(2.1) we find that $(3,2)_{1/6}$ ($Q=1/6$) in the untwisted sector
 should satisfy
$$ {7\over 12}+{1\over 36k_1}+h'-1=0, \eqno{(2.5)} $$
where $h'$ represents extra U(1) contributions to $h$.
Since $h'\ge 0$, we have a restriction $k_1\geq 1/15$ to exist $(3,2)_{1/6}$
 in the untwisted sector.
Similary, we get restrictions for the other representations in the untwisted
 sector, as shown in Table 2.

{}From eq.~(2.2) we find that $(3,2)_{1/6}$ in the twisted sector $T_{\ell m}$
 has oscillators with $N_{\rm OSC}\le 5/12-c_{\ell m}$ under a condition
 $k_1 \ge 1/(15-36N_{\rm OSC}-36c_{\ell m})$.
Similarly we can obtain the allowable values of $N_{\rm OSC}$ for the other
 representations.

\vspace{0.8 cm}
\leftline{\large \bf 3. Duality and Threshold Corrections}
\vspace{0.8 cm}

It is plausible that the duality symmmetry is retained in effective field
 theories derived from the orbifold models.
In the theories, moduli fields $T_i$ ($i=1,2,3$) associated with the $i$-th
 complex planes have the K\"ahler potentials
$$ -\sum_i{\rm log}|T_i+\overline T_i|,\eqno(3.1)$$
which are invariant under a duality transformation:
$$ T_i \rightarrow {a_iT_i-ib_i \over ic_iT_i+d_i} ,\eqno(3.2)$$
up to the K\"ahler transformation, where $a_i,b_i,c_i,d_i\in {\bf Z}$ and
 $a_id_i-b_ic_i=1$.
The K\"ahler potential of the matter field $A$ is
$$ \prod^3_{i=1}(T_i+\overline T_i)^{n^i}A\overline A,\eqno(3.3)$$
whose duality invariance requires the following transformation:
$$ A \rightarrow A \prod_{i=1}^3(ic_iT_i+d_i)^{n^i},\eqno(3.4)$$
where $n^i$ is called a modular weight \cite{Dixon2,Ibanez}.
Hereafter, we consider a case of an overall modulus $T_1=T_2=T_3=T$,
 for simplicity.
The sum of the modular weight elements $n=\sum_i n^i$ is available in this
 case.

For the untwisted sector, the matter fields have $n=-1$.
The twisted sector $T_{\ell m}$ has the following modular weights:
$$\begin{array}{llll}
n&=&-1-p+q&\quad {\rm for\ the\ twisted\ sectors\ with\ some\ unrotated\
 planes}, \\
n&=&-2-p+q&\quad {\rm for\ the\ twisted\ sectors\ without\ any\ unrotated\
 plane},
\end{array} $$
where $p$ ($q$) is a number of the oscillators $\partial X_i$
 ($\partial \tilde X_i$) contributing to the massless state in the twisted
 sector.
Table 3 and Table 4 list the values of $n$ and lower bounds of $k_1$ permitted
 by the massless condition in the previous section.
For example, twisted sectors $T_{01}$ and $T_{05}$ of $Z_3\times Z_6$ in the
 fourth column and the third row of Table 4 are possible to have (3,2) with
 $n=0$ , $-1$ and $-2$, if $k_1\geq 1/4$, $1/10$ and $1/4$ respectively.

The threshold corrections of the gauge coupling constants are induced by the
 tower of higher massive modes and depend on the overall modulus $T$.
They are given by
$$\Delta_a(T)=-{1 \over 16\pi^2}(b'_a-k_a\delta_{\rm GS}){\rm log}|\eta(T)|^4,
 \eqno(3.5)$$
where $\delta_{\rm GS}$ is a Green-Schwarz coefficient \cite{GS} independent
 of the gauge groups $a\ (=3,2$ and 1 correspond to SU(3), SU(2) and U(1)
 respectively) and $\eta(T)=e^{-\pi T/12}\prod_{n \ge 1}(1-e^{-2\pi n T})$ is
 the Dedekind function \cite{Derendinger,Dixon}.
Anomaly coefficients $b'_a$  are obtained from
$$b'_a=-3C(G_a)+\sum_{\underline R} T(\underline R)\left(3+2n_{\underline R}
 \right),\eqno(3.6)$$
where $n_{\underline R}$ is an overall modular weight of a representation
 $\underline R$ and $T(\underline R)$ is an index given by
$T(\underline R)=C(\underline R){\rm dim}(\underline R)/{\rm dim}(G)$, e.g.,
$T(\underline R)=1/2$ for the fundamental representation of SU($N$).

Using the threshold correction, we obtain the one-loop coupling constants
 $\alpha_a(\mu)=k_ag_a^2(\mu)/4\pi$ at an energy scale $\mu$ as follows,
$$ \alpha^{-1}_a(\mu)=\alpha^{-1}_{\rm string}+{1 \over 4\pi}{b_a\over k_a}
 {\rm log}{M_{\rm string}^2 \over \mu^2}-{1\over 4\pi}
 ({b_a^{\prime }\over k_a}- \delta_{\rm GS}){\rm log}[(T+\overline{T})
 |\eta(T)|^4],\eqno(3.7)$$
where $\alpha_{\rm string}=g^2_{\rm string}/4\pi$ and $b_a$ are $N=1$
 $\beta$-function coefficients.
We get the same $b_3=-3$, $b_2=1$ and $b_1=11$ as ones of the MSSM, because we
 consider the minimal string model after the removal of the extra pairs of
 $(3,2)+(\bar 3,2)$, etc. and the extra U(1)'s.
As discussed in the previous section, the level $k_1$ is the arbitrary
 constant, while $k_3=k_2=1$.

{}From this renormalization group flow, we can derive a unified scale $M_X$
from
$$ {\rm log}{M_{X} \over M_{\rm string}}={b_{3}^{\prime }-
b_{2}^{\prime } \over 8}{\rm log}[(T+\overline{T})|\eta(T)|^4] ,\eqno(3.8)$$
where  $\alpha^{-1}_3(M_X)=\alpha^{-1}_2(M_X)$.\footnote{When $k_1=5/3$,
 the LEP measurements are consistent with $\alpha^{-1}_3(M_X)=\alpha^{-1}_2
 (M_X)=\alpha^{-1}_1(M_X)$ within the framework of the MSSM.}
Since the LEP measurements indicate $M_X<M_{\rm string}$, eq.~(3.8) gives a
 restriction
$$\Delta b'\equiv b'_3-b'_2 > 0,\eqno(3.9)$$
because log$[(T+\overline{T})|\eta(T)|^4]$ is always negative.
{}From eq.~(3.6), we get
$$ \Delta b'=-6 -\sum n_{(3,2)}+\sum n_{(\bar 3,1)}-\sum n_{(1,2)}
 \eqno{(3.10)} $$
It has been pointed out \cite{Ibanez} that any $Z_2\times Z_2$ orbifold model
 does not satisfy the condition (3.9), because $\Delta b'=-4$ as derived from
 eq.~(3.10), Table 2 and Table 3.
When $\Delta b'=3$, we get $T\simeq 12$ through $M_X\simeq 10^{16}$ GeV and
 $M_{\rm string}\simeq 3.73\times 10^{17}$ GeV.
Since such a large $T$ is not desired, we impose $\Delta b'>3$ on the
 subsequent analyses.

The one-loop fine structure constant of the electro-magnetic interaction is
 obtained from $\alpha^{-1}_{\rm em}=k_1\alpha^{-1}_1+\alpha^{-1}_2$ as
$$ \alpha^{-1}_{\rm em}(\mu)=(k_1+1)\alpha^{-1}_{\rm string}+{3 \over
 \pi}{\rm log}{M_{\rm string}^2 \over \mu^2}$$
$$ \qquad\qquad\qquad -{1\over 4\pi}(b_1^{\prime}+b_2^{\prime}-(k_1+1)
 \delta_{\rm GS}){\rm log}[(T+\overline{T})|\eta(T)|^4].\eqno(3.11)$$
By means of $\sin ^2\theta_{\rm W}=\alpha_{\rm em}/\alpha_2$, we derive
$$ \sin ^2\theta_{\rm W}(\mu)={1\over k_1+1}+{\alpha_{\rm em}(\mu)\over 4\pi
 (k_1+1)}\big( (k_1-11){\rm log}{M_{\rm string}^2 \over \mu^2}$$
$$ \qquad\qquad\qquad -(k_1b_2^{\prime}-k_1b_2^{\prime}){\rm log}[(T+
 \overline{T})|\eta(T)|^4]\big) .\eqno(3.12)$$

\vspace{0.8 cm}
\leftline{\large \bf 4. Level $k_1$ and Overall Modulus $T$}
\vspace{0.8 cm}

At first, we discuss the level $k_1$ of the U(1)$_Y$.
Using eqs.~(3.8), (3.11) and (3.12), we obtain
$$k_1={12\Delta b'{\rm log }(M_{\rm sting}^2 / \mu^2)-4(b'_1+b'_2){\rm log}
 (M_X^2 / M_{\rm string}^2)- 4 \pi\Delta b'\alpha^{-1}_{\rm em}(\mu) \over
 \Delta b'{\rm log }(M_{\rm sting}^2 / \mu^2)-4b'_2{\rm log}
 (M_X^2 / M_{\rm string}^2)-4 \pi\Delta b'\alpha^{-1}_{\rm em}(\mu)
 {\rm sin}^2\theta_W(\mu)}-1,  \eqno{(4.1)} $$
where $\mu$ is higher than the SUSY breaking scale.
The unified scale $M_X$ is derived from
$$ {\rm log}(M_X^2/ \mu^2)=\pi \{\sin^2\theta_{\rm W}(\mu)\alpha^{-1}_{\rm em}
 (\mu)-\alpha_3^{-1}(\mu)\}. \eqno(4.2) $$
The experimental values are $\sin ^2\theta_{\rm W}(M_Z)=0.2325\pm 0.0008$,
 $\alpha_{\rm em}^{-1}(M_Z)=127.9$ and $\alpha^{-1}_3(M_Z)=8.82\pm 0.27$ at
 $M_Z=91.173\pm 0.020$ GeV.
Supposing that the SUSY scale is $M_Z$, we derive $M_X=10^{16.22\pm 0.25}$
 GeV.

The anomaly coefficients $b_a^{\prime}$ are obtained from inserting values of
 three $n_{(3,2)}$, three $n_{(\bar 3,1)_{1/3}}$, three
 $n_{(\bar 3,1)_{-2/3}}$, five $n_{(2,1)}$ and three $n_{(1,1)}$ permitted in
 Table 2, Table 3 and Table 4 into eq.~(3.6).
The values of them and the experimental values give the level $k_1$ through
 eq.~(4.1).
The modular weights $n_{\underline R}$ of each representations contributing
 to $b_a^{\prime}$ must be checked whether the obtained $k_1$ is not smaller
 than the lower bounds imposing on the levels for each $n_{\underline R}$.
In this way, we have found no permissible combination of $n_{\underline R}$ in
 $Z_3\times Z_3$, but 400 in $Z_2\times Z_4$, 78957 in $Z_2\times
 Z_6^{\prime}$, 29000 in $Z_2\times Z_6$ and 198136 in $Z_6\times Z_6$ through
 computer analyses.
Here $Z_2\times Z_4$ and $Z_4\times Z_4$ ($Z_3\times Z_6$ and $Z_6\times Z_6$)
 have the same permissible combinations of $n_{\underline R}$ as found in
 Table 3 and Table 4.
It follows that we treat them as the identical orbifolds.
The fourth column of Table 1 lists the minimum values and the maximum values
 of $k_1$ among the permissible combinations.
These values include 17\%, 25\%, 21\% and 23\% experimental errors for $Z_2
 \times Z_4$, $Z_2\times Z_6^{\prime}$, $Z_2\times Z_6$ and $Z_3\times Z_6$,
 respectively.
The lower bound $k_1\geq 1.00$ is given by the singlets as shown in Table 2,
 Table 3 and Table 4.

When the SUSY scale is 1 TeV, we get $\alpha_{\rm em}^{-1}({\rm 1\ TeV}) =
 127.2$, $\sin^2\theta _{\rm W}({\rm 1\ TeV})=0.2432\pm 0.0021$ and
 $\alpha^{-1}_3({\rm 1\ TeV})=11.48\pm 0.27$ from the non-SUSY renomalization
 group calculations.
Using these values and estimating similarly as the case of the $M_Z$ SUSY
 scale, we get the ranges of $k_1$ as listed in the fifth column of Table 1,
 which include 12\%, 17\%, 25\%, 21\% and 22\% experimental errors in $Z_3
 \times Z_3$, $Z_2\times Z_4$, $Z_2\times Z_6^{\prime}$, $Z_2\times Z_6$ and
 $Z_6\times Z_6$, respectively.
In this case, six combinations for $Z_3\times Z_3$ are permitted.
One of them, for example, is given by $n_{(3,2)}=-2$, $n_{(\bar 3,1)_{1/3}}=
 0$, $n_{(\bar 3,1)_{-2/3}}=-2$, $n_{(1,2)}=-2$ and $n_{(1,1)}=-1$ for each
 representations and it gives $k_1=1.21\pm 0.14$, i.e., the maximum value in
 $Z_3\times Z_3$.
It is also found that the other orbifolds $Z_2\times Z_4$, $Z_2\times
 Z_6^{\prime}$, $Z_2\times Z_6$ and $Z_3\times Z_6$ have permissible
 combinations of 840, 91182, 38949 and 279618 respectively.

{}From Table 1, we notice that the ranges of $k_1$ in the $M_Z$ SUSY scale are
 almost same as ones in the 1 TeV SUSY scale except for $Z_3\times Z_3$.
It is also noticed that the GUT prediction $k_1=5/3$ is not included in
 $Z_3\times Z_3$ and $Z_2\times Z_4$.

The ranges of $k_1$ forbid some modular weights of the matter fields in
 Table 3 and Table 4.
In particular the modular weights of $(\bar 3,1)_{-2/3}$ and $(1,1)_1$ are
 restricted tightly.
In $Z_3\times Z_6$, for instance, $(\bar 3,1)_{-2/3}$ is not prohibitted to
 possess $n=2,1,-3,-4$ on $T_{01},T_{05}$, $n=0,-2$ on $T_{02},T_{04},T_{10},
 T_{14},T_{20},T_{22}$, $n=-3,-4$ on $T_{11},T_{13}$ and $n=-1,-4$ on $T_{21}$
 for the unrestricted $k_1$, but for $k_1\leq 2.1$.
These restrictions of the modular weights rule out higher $N_{\rm OSC}$.
Thus, one can reduce extents of the minimal string model searches on the
 $Z_N\times Z_M$ orbifolds.

Next, we investigate minimum values of the overall modulus $T$.
{}From eq.~(3.8) we can get Re$T$ when $\Delta b'$ is obtained.
If $k_1$ is not restricted at all, the maximum values of $\Delta b'$ are
 easily obtained from Table 2, Table 3 and Table 4 through eq.~(3.10).
For example, $Z_3\times Z_6$ gives $\Delta b'\leq 32$ which leads to
 $T\geq 2.4$ from eq.~(3.8).
Lower bounds of the overall moduli $T_{\rm Low}$ of the other orbifolds are
 found in the third column of Table 1.

These lower bounds $T_{\rm Low}$ may be heightened when $k_1$ are restricted
 as discussed above.
In $Z_3\times Z_6$, for example, both $k_1\leq 2.11$ at the $M_Z$ SUSY scale
 and $k_1\leq 2.09$ at the 1 TeV SUSY scale give $\Delta b'\leq 26$.
This leads to a minimum value of the overall modulus $T_{\rm Min}=2.7$.
The sixth column of Table 1 lists $T_{\rm Min}$ of the whole orbifolds, where
 $k_1$ at the $M_Z$ and 1 TeV SUSY scales give the identical $T_{\rm Min}$ of
 6.4, 4.4, 4.0, 3.4 for $Z_3\times Z_3$, $Z_2\times Z_4$, $Z_2\times
 Z_6^{\rm \prime}$, $Z_2\times Z_6$, respectively.
In $Z_2\times Z_6^{\prime}$, $T_{\rm Min}$ is equal to $T_{\rm Low}$ because
 $\Delta b'\leq 14$ for the unrestricted $k_1$ is not changed for $k\leq 2.0$.
It is noted that $Z_3\times Z_6$ may derive the smallest value of $T$.

\vspace{0.8 cm}
\leftline{\large \bf 5. Conclusions}
\vspace{0.8 cm}

In this paper we have discussed the minimal string model derived from the
 $Z_N\times Z_M$ orbifold models.
In the case of the overall moduli, we have studied the threshold corrections
 of the gauge coupling constants.
Using the computer analyses, we have investigated the ranges of the levels
 $k_1$ of U(1)$_Y$ consistent with the experimental values of $\sin^2
 \theta_{\rm W}$, $\alpha^{-1}_{\rm em}$, $\alpha^{-1}_3$ and $M_Z$.
It has been found that the $Z_3\times Z_3$ model with the $M_Z$ SUSY breaking
 scale is ruled out due to nonexistence of the allowable values of $k_1$.
The GUT prediction $k_1=5/3$ is not included in the ranges $1.06\leq k_1\leq
 1.35$ in $Z_3\times Z_3$ at the 1 TeV SUSY scale and $1.00\leq k_1\leq 1.60$
 ($1.00\leq k_1\leq 1.61$) in $Z_2\times Z_4$ and $Z_4\times Z_4$ at the $M_Z$
 (1 TeV) SUSY scale.
The minimum values of the overall moduli have been obtained for explaining the
 discrepancy between the unified scale $M_X$ of $g_3$ and $g_2$ and the string
 scale $M_{\rm string}$.
We have found that the $Z_3\times Z_3$, $Z_2\times Z_4$ ($Z_4\times Z_4$),
 $Z_2\times Z_6^{\rm \prime}$, $Z_2\times Z_6$ and $Z_3\times Z_6$
 ($Z_6\times Z_6$) models have $T\geq 6.4$, 4.4, 4.0, 3.4 and 2.7,
 respectively.

Although we have considered the overall modulus in this paper, the above
 procedure can be also applied to cases without such a restriction.
One will be able to investigate the minimal string model with extra matters
 \cite{extra} through the similar estimations.
It is also interesting to consider the level $k_1$ in $Z_N$ orbifold models.

\vspace{0.8 cm}
\newpage
\leftline{\large \bf Acknowledgement}
\vspace{0.8 cm}

The authors would like to thank D.~Suematsu for useful discussions.


\newpage
\pagestyle{empty}
\noindent
\begin{center}
{\bf \large Table 1. Restrictive values of $k$ and $T$}

\vspace{10mm}

\begin{tabular}{|c|cc||c|c|c|c|}
\hline
Orbifold           & $v_1$      & $v_2$      &
$T_{\rm Low}$ & $k_1(M_Z)$   & $k_1$(1 TeV)   & $T_{\rm Min}$ \\ \hline \hline
$Z_2\times Z_2$    & (1,0,1)/2  &  (0,1,1)/2 &
     -     &   -           &       -          &    -     \\ \hline
$Z_3\times Z_3$    & (1,0,2)/3  &  (0,1,2)/3 &
    3.8    &   -           & $1.06\sim 1.35$  &   6.4      \\ \hline
$Z_2\times Z_4$    & (1,0,1)/2  &  (0,1,3)/4 &
    3.8    & $1.00\sim 1.60$ & $1.00\sim 1.61$&   4.4      \\ \hline
$Z_4\times Z_4$    & (1,0,3)/4  &  (0,1,3)/4 &
  \multicolumn{4}{c|}{same as $Z_2\times Z_4$}          \\ \hline
$Z_2\times {Z_6}'$ & (1,0,1)/2  &  (1,1,4)/6 &
    4.0    & $1.00\sim 2.08$ & $1.00\sim 2.06$&   4.0   \\ \hline
$Z_2\times Z_6$    & (1,0,1)/2  &  (0,1,5)/6 &
    2.4    & $1.00\sim 1.98$ & $1.00\sim 1.96$&   3.4      \\ \hline
$Z_3\times Z_6$    & (1,0,2)/3  &  (0,1,5)/6 &
    2.4    & $1.00\sim 2.11$ & $1.00\sim 2.09$&   2.7   \\ \hline
$Z_6\times Z_6$    & (1,0,5)/6  &  (0,1,5)/6 &
  \multicolumn{4}{c|}{same as $Z_3\times Z_6$}          \\ \hline
                                   \end{tabular}

\end{center}
\vspace{20mm}
\begin{center}
{\bf \large Table 2. Lower-bound of $k_1$ in untwisted sector}
\vspace{10mm}

\begin{tabular}{|c|c|c|c|c|c|}
\hline
        & \multicolumn{5}{c|}{Lower-bound of $k_1$} \\ \hline
  $n$   & $(3,2)$ & $(\bar 3,1)_{1/3}$ & $(\bar 3,1)_{-2/3}$ & $(1,2)$ &
 $(1,1)$  \\ \hline \hline
  $-1$  &   1/15  &      1/6      &      2/3      &   1/3   & 1   \\ \hline
                                   \end{tabular}

\end{center}
\newpage

\noindent

\begin{center}
{\bf \large Table 3. Lower-bound of $k_1$ in twisted sectors (I)}
\vspace{10mm}

\begin{tabular}{|c|c|c||r|c|c|c|c|c|c|}
\hline
 \multicolumn{3}{|c||}{T-sec.}
        &     & \multicolumn{5}{c|}{Lower-bound of $k_1$} \\ \hline
 $Z_2\times Z_2$  & $Z_2\times Z_4$ &  $Z_4\times Z_4$
        & $n$ & $(3,2)$ & $(\bar 3,1)_{1/3}$ & $(\bar 3,1)_{-2/3}$ & $(1,2)$
 & $(1,1)$  \\ \hline \hline
         -        & $T_{01},T_{03}$ & $T_{01}$,$T_{03}$,
        &  2  &    -    &       -       &      -        &   -     & 16  \\
                  &                 & $T_{10}$,$T_{13}$,
        &  1  &    -    &       -       &      -        &    4    & 16/5  \\
                  &                 & $T_{30}$,$T_{31}$
        &  0  &    -    &    16/33      &    64/33      &    4/5  & 16/9  \\
                  &                 &
        &$-1$ &  4/33   &    16/69      &    64/69      &    4/9  & 16/13 \\
                  &                 &
        &$-2$ &    -    &    16/33      &    64/33      &    4/5  & 16/9  \\
                  &                 &
        &$-3$ &    -    &       -       &      -        &    4    & 16/5  \\
                  &                 &
        &$-4$ &    -    &       -       &      -        &    -    & 16    \\
\hline
$T_{01},T_{10},$  & $T_{02},T_{10}$,& $T_{02}$,$T_{20}$,
        &  0  &    -    &       -       &      -        &    -    & 4     \\
$T_{11}$          & $T_{12}$        & $T_{22}$
        &$-1$ &  1/6    &    4/15       &    16/15      &    1/2  & 4/3   \\
                  &                 &
        &$-2$ &    -    &       -       &      -        &    -    & 4     \\
\hline
     -            & $T_{11}$        & $T_{11}$,$T_{12}$,
        &$-1$ &    -    &       -       &      -        &    -    & 16/3  \\
                  &                 &  $T_{21}$
        &$-2$ &  4/15   &   16/51       &    64/51      &    4/7  & 16/11 \\
                  &                 &
        &$-3$ &    -    &   16/15       &    64/15      &    4/3  & 16/7  \\
                  &                 &
        &$-4$ &    -    &       -       &      -        &    -    & 16/3  \\
\hline
                                   \end{tabular}

\newpage

{\bf \large Table 4. Lower-bound of $k_1$ in twisted sectors (II)}

\vspace{10mm}
\footnotesize
\begin{tabular}{|c|c|c|c|c||r|c|c|c|c|c|c|}
\hline
 \multicolumn{5}{|c||}{T-Sec.} &     & \multicolumn{5}{c|}{Lower-bound of
 $k_1$}  \\ \hline
 $Z_3\times Z_3$  & $Z_2\times {Z_6}'$ &  $Z_2\times Z_6$ &  $Z_3\times Z_6$
 &  $Z_6\times Z_6$
  & $n$ & $(3,2)$ & $(\bar 3,1)_{1/3}$ & $(\bar 3,1)_{-2/3}$ & $(1,2)$ &
 $(1,1)$  \\ \hline \hline
        -         &        -           & $T_{01},T_{05}$  &
 $T_{01},T_{05}$  &  $T_{01}$,$T_{05}$,
  &  4  &    -    &       -       &      -        &   -     & 36     \\
                  &                    &                  &
                  &  $T_{10}$,$T_{15}$,
  &  3  &    -    &       -       &      -        &   -     & 36/7  \\
                  &                    &                  &
                  & $T_{50}$,$T_{51}$
  &  2  &    -    &      4        &      16       &    9/4  & 36/13 \\
                  &                    &                  &
                  &
  &  1  &    -    &     4/7       &    16/7       &    9/10 & 36/19 \\
                  &                    &                  &
                  &
  &  0  &   1/4   &     4/13      &    16/13      &    9/16 & 36/25 \\
                  &                    &                  &
                  &
  &$-1$ &  1/10   &     4/19      &    16/19      &    9/22 & 36/31 \\
                  &                    &                  &
                  &
  &$-2$ &   1/4   &     4/13      &    16/13      &    9/16 & 36/25 \\
                  &                    &                  &
                  &
  &$-3$ &    -    &     4/7       &    16/7       &    9/10 & 36/19 \\
                  &                    &                  &
                  &
  &$-4$ &    -    &      4        &      16       &    9/4  & 36/13 \\
                  &                    &                  &
                  &
  &$-5$ &    -    &       -       &      -        &   -     & 36/7  \\
                  &                    &                  &
                  &
  &$-6$ &    -    &       -       &      -        &    -    & 36    \\ \hline
$T_{01},T_{02}$,  &         -          & $T_{02},T_{04}$  &
$T_{02},T_{04}$,  & $T_{02}$,$T_{04}$,
  &  1  &    -    &       -       &      -        &    -    & 9     \\
$T_{10},T_{12}$,  &                    &                  &
$T_{10},T_{14}$,  & $T_{20}$,$T_{24}$,
  &  0  &    -    &      1        &      4        &    9/7  & 9/4   \\
$T_{20},T_{21}$   &                    &                  &
$T_{20},T_{22}$   & $T_{40}$,$T_{42}$
  &$-1$ &  1/7    &    1/4        &      1        &    9/19 & 9/7   \\
                  &                    &                  &
                  &
  &$-2$ &    -    &      1        &      4        &    9/7  & 9/4   \\
                  &                    &                  &
                  &
  &$-3$ &    -    &       -       &      -        &    -    & 9     \\ \hline
       -          & $T_{03},T_{10}$,   & $T_{03},T_{10}$, &
$T_{03}$          & $T_{03}$,$T_{30}$,
  &  0  &    -    &       -       &      -        &    -    & 4     \\
                  & $T_{13}$           & $T_{13}$         &
                  &    $T_{33}$
  &$-1$ &  1/6    &    4/15       &    16/15      &    1/2  & 4/3   \\
                  &                    &                  &
                  &
  &$-2$ &    -    &       -       &      -        &    -    & 4     \\ \hline
       -          &          -         & $T_{11},T_{12}$, &
$T_{11},T_{13}$   & $T_{12}$,$T_{13}$,
  &$-1$ &    -    &       -       &      -        &    -    & 36/7  \\
                  &                    &                  &
                  & $T_{21}$,$T_{23}$,
  &$-2$ &  1/4    &    4/13       &     16/13     &    9/16 & 36/25   \\
                  &                    &                  &
                  & $T_{31}$,$T_{32}$
  &$-3$ &    -    &    4/7        &     16/7      &    9/10 & 36/19   \\
                  &                    &                  &
                  &
  &$-4$ &    -    &      4        &      16       &    9/4  & 36/13   \\
                  &                    &                  &
                  &
  &$-5$ &    -    &       -       &      -        &    -    & 36/7    \\
                  &                    &                  &
                  &
  &$-6$ &    -    &       -       &      -        &    -    & 36     \\ \hline
 $T_{11}$         & $T_{02}$           &         -        &
 $T_{12}$         & $T_{22}$
  &$-2$ &  1/3    &    1/3        &     4/3       &    3/5  & 3/2   \\
                  &                    &                  &
                  &
  &$-3$ &    -    &       -       &      -        &    3    & 3     \\ \hline
       -          & $T_{01},T_{11}$,   &         -        &
 $T_{21}$         & $T_{11}$,$T_{14}$,
  &  0  &    -    &       -       &      -        &   -     & 12     \\
                  & $T_{14}$           &                  &
                  & $T_{41}$
  &$-1$ &    -    &     4/3       &    16/3       &    3/2  & 12/5   \\
                  &                    &                  &
                  &
  &$-2$ &  1/6    &     4/15      &    16/15      &    1/2  & 4/3    \\
                  &                    &                  &
                  &
  &$-3$ &    -    &     4/9       &    16/9       &    3/4  & 12/5   \\
                  &                    &                  &
                  &
  &$-4$ &    -    &     4/3       &    16/3       &    3/2  & 12/5   \\
                  &                    &                  &
                  &
  &$-5$ &    -    &       -       &      -        &   -     & 4      \\
                  &                    &                  &
                  &
  &$-6$ &    -    &       -       &      -        &    -    & 12     \\ \hline
                                   \end{tabular}

\end{center}
\end{document}